\documentclass[prl,twocolumn,showpacs,preprintnumbers,amsmath,amssymb, superscriptaddress]{revtex4-2}

\usepackage[makeroom]{cancel}

\usepackage{amsmath}    % need for subequations
\usepackage{amssymb,environ}
\usepackage{graphicx}   % need for figures
\usepackage{verbatim}   % useful for program listings
\usepackage{color}      % use if color is used in text
\usepackage{hyperref}   % use for hypertext links, including those to external documents and URLs
\usepackage[normalem]{ulem}
\usepackage{natbib}
\usepackage{enumitem}

\hypersetup{colorlinks,linkcolor=blue,urlcolor=blue,citecolor=blue}

\definecolor{darkgreen}{rgb}{0,0.5,0}
\definecolor{purple}{rgb}{0.35,0,0.35}
\definecolor{orange}{rgb}{1,0.5,0}
\definecolor{darkred}{rgb}{.7,0,0}
\definecolor{darkblue}{rgb}{0,0,.3}
\definecolor{grey}{rgb}{.6,.6,.6}
\definecolor{dimgreen}{rgb}{0.2,0.6,0.1}

\begin{document}

\newcommand{\kv}[0]{\mathbf{k}}
\newcommand{\Rv}[0]{\mathbf{R}}
\newcommand{\Hv}[0]{\mathbf{H}}
\newcommand{\Mv}[0]{\mathbf{M}}
\newcommand{\Vv}[0]{\mathbf{V}}
\newcommand{\Uv}[0]{\mathbf{U}}
\newcommand{\rv}[0]{\mathbf{r}}
\newcommand{\gv}[0]{\mathbf{g}}
\newcommand{\al}[0]{\mathbf{a_{1}}}
\newcommand{\as}[0]{\mathbf{a_{2}}}
\newcommand{\K}[0]{\mathbf{K}}
\newcommand{\Kp}[0]{\mathbf{K'}}
\newcommand{\dkv}[0]{\delta\kv}
\newcommand{\dkx}[0]{\delta k_{x}}
\newcommand{\dky}[0]{\delta k_{y}}
\newcommand{\dk}[0]{\delta k}
\newcommand{\cv}[0]{\mathbf{c}}
\newcommand{\dv}[0]{\mathbf{d}}
\newcommand{\qv}[0]{\mathbf{q}}
\newcommand{\Tr}[0]{\mathrm{Tr}}
\newcommand{\Gv}[0]{\mathbf{G}}
\newcommand{\ev}[0]{\mathbf{e}}
\newcommand{\uu}[1]{\underline{\underline{#1}}}
\newcommand{\ket}[1]{|#1\rangle}
\newcommand{\bra}[1]{\langle#1|}
\newcommand{\average}[1]{\langle #1 \rangle}
\newcommand{\scrap}[1]{{\color{orange}{\sout{#1}}}}
\newcommand{\red}[1]{{\color{red}#1}}
\newcommand{\be}{\begin{equation}}
\newcommand{\ee}{\end{equation}}
\newcommand{\kB}[0]{k_\mathrm{B}}

\newcommand{\jav}[1]{{\color{red}#1}}

%\title{Thermalization after quantum and dissipative quench in Luttinger liquid}
\title{Poor man's approach to thermalization in Luttinger liquid}
%\title{The Lindbladian route towards thermalization of a Luttinger liquid}

\author{\'Ad\'am B\'acsi}
\email{bacsi.adam@sze.hu}
\affiliation{MTA-BME Lend\"ulet Topology and Correlation Research Group,
Budapest University of Technology and Economics,  M\H uegyetem rkp. 3., H-1111 Budapest, Hungary}
\affiliation{Department of Mathematics and Computational Sciences, Sz\'echenyi Istv\'an University, 9026 Gy\H or, Hungary}
\author{Bal\'azs D\'ora}
\affiliation{MTA-BME Lend\"ulet Topology and Correlation Research Group,
Budapest University of Technology and Economics,  M\H uegyetem rkp. 3., H-1111 Budapest, Hungary}
\affiliation{Department of Theoretical Physics, Institute of Physics, Budapest University of Technology and Economics, M\H uegyetem rkp. 3., H-1111 Budapest, Hungary}
\date{\today}

\begin{abstract}
We study the non-equilibrium dynamics of a Luttinger liquid after a simultaneous quantum quench of the interaction 
and dissipative quench to the environment within the realm of the Lindblad equation. 
When the couplings to environment satisfy detailed balance, the system is destined to thermalize, which we follow using bosonization.
The thermodynamic entropy of the system, which also encodes information about the entanglement with the environment, either exhibits a maximum and minimum or grows 
monotonically before reaching its thermal value in a universal fashion.
%This applies to other thermodynamics quantities such as the total energy or the R\'enyi entropy.
The single-particle density matrix reveals rich behaviour in the spatio-temporal "phase diagram", including thermal, prethermal and dissipation enhanced sudden quench Luttinger liquid behaviour as well as dissipation enhanced Fermi liquid response. 
%Our approach could be applied to a large variety of systems.
\end{abstract}

\maketitle

\paragraph{Introduction.}
Non-equilibrium dynamics and quantum quenches have attracted enormous attention ever since the manipulation of quantum systems became experimentally accessible in 
cold atomic gases\cite{polkovnikov2011_revmod,blochRMP2008,dziarmagareview,cazalillarmp}. 
In particular, sophisticated experimental methods enable the study of non-equilibrium dynamics in one-dimensional quantum systems \cite{erne,gring}.
These studies provide essential information about relaxation and equilibration, prethermalization and thermalization in a large variety of settings,
incorporating topological and strongly interacting systems as well.
Understanding non-equilibrium quantum dynamics is also essential for applications in quantum metrology\cite{toth}, quantum computation and information 
processing\cite{nielsen}.

Among the most fascinating topics, thermalization represents a central theme in non-equilibrium dynamics\cite{kinoshita,mori,kaminishinatphys}.
Closed quantum systems starting from a pure state do not thermalize due to unitary dynamics. However, a small subsystem can in principle thermalize since
the rest of the system can act as an effective heat bath. This phenomenon is absent from integrable systems, where the large number of constants of motion
prevents the subsystem from thermalization\cite{deutsch,srednickipre}. Non-integrable systems, on the other hand, display subsystem 
thermalization\cite{rigol}, though these are rather hard to analyze\cite{bertini} due to their non-integrability.

Here we take a different approach to study thermalization and investigate open quantum systems. In particular, we focus on strongly interacting one-dimensional fermions, 
undergoing
a sudden quantum quench\cite{widera,karrasch,cazalillaprl,perfettoEPL,gutman,buchhold2016,ruggeiro2021,kaminishi,moosavi} as well as dissipative coupling to environment\cite{breuer,daley}. The 
resulting open quantum system thermalizes when couplings to environment satisfy detailed 
balance\cite{rajagopal,breuer,ashidathermal}. As a result, we are able to follow \emph{exactly} the build up of Luttinger liquid (LL) physics and its competition with non-unitary dynamics due to 
environment and the eventual thermalization of the system.

We find the the thermodynamic entropy of the system, which quantifies the entanglement with environment, can follow two qualitatively distinct
time evolutions, depending on the environment temperature and interaction strength. It either exhibits a maximum, then a minimum or
grows monotonically before reaching the thermal value. We argue that these features are typical in thermalizing open quantum systems.
The single particle density matrix reveals a whole zoo of strongly correlated behaviours, such as
dissipation enhanced sudden quench LL and Fermi liquid behaviour, thermal and prethermal LL.
Due to the generality and universality of the LL picture, our approach applies to a great variety of low dimensional fermionic, bosonic and spin systems.

\paragraph{Thermalizing Luttinger liquid.}
We study the low-energy behavior of a one-dimensional fermionic system, forming a Luttinger liquid\cite{giamarchi}. 
The system is prepared in the ground state of a non-interacting Fermi gas, and the fermionic interaction is switched on suddenly at $t=0$. 
Upon bosonization, the time-dependent Hamiltonian is 
\begin{gather}
H=\sum_{q>0}\left[\omega_q(t)\left(b_q^+b_q + b_{-q}^+ b_{-q}\right) + g_q(t) \left(b_q b_{-q} + b_q^+ b_{-q}^+\right)\right],
\label{eq:ham}
\end{gather}
where $\omega_q(t)=v_F q + \Theta(t) g_4 q$ and $g_q(t)=\Theta(t) g_2 q$ with $v_F$ the Fermi velocity and $g_2$ and $g_4$ describing the forward scattering induced by the interaction between fermions of opposite and same chiralities, respectively \cite{giamarchi}. 
%The time dependence of $\omega_q(t)$ and $g_q(t)$ indicate that for $t<0$, the system is described by a non-interacting Hamiltonian and the interaction is suddenly switched on at $t=0$.
For $t>0$, the interacting Hamiltonian is diagonalized by a Bogoliubov transformation \cite{giamarchi} as
\begin{gather}
H = E_0 + \sum_{q>0} c q \left(d_q^+ d_q + d_{-q}^+ d_{-q}\right),
\end{gather}
where $c=\sqrt{(v_F+g_4)^2-g_2^2}$ is the renormalized sound velocity and $E_0 = \sum_{q>0} (c-v_F-g_4)q$ is the ground state energy.
The sudden quantum quench described in Eq. \eqref{eq:ham} has been extensively investigated in previous studies 
\cite{Cazalilla2016,cazalillaprl,Iucci_Cazalilla_LLquench_PRA2009}. 

In addition to sudden quantum quench, we also couple the system to environment at $t=0$, therefore dissipative processes are also taken into account
during the time evolution. For simplicity, we use the Lindblad equation\cite{breuer} for the 
density matrix as $\partial_t \rho = -i\left[H,\rho\right] + \mathcal{D}[\rho(t)]$ where the dissipation reads 
as
\begin{gather}
\mathcal{D}[\rho(t)] = \sum_{q} \gamma |q|\Big[(n(T,q) + 1) \mathcal{D}_{\downarrow,q}[\rho] +  n(T,q)\mathcal{D}_{\uparrow,q}[\rho]\Big]
\label{eq:jumps}
\end{gather}
involving $\mathcal{D}_{\downarrow,q}[\rho] = d_{q}\rho d_q^+  - \frac{1}{2}\left\{d_q^+ d_q,\rho\right\}$ and $\mathcal{D}_{\uparrow,q}[\rho] = d_{q}^+\rho d_q  - \frac{1}{2}\left\{d_q d_q^+,\rho\right\}$.
This approach remain meaningful in the weak coupling, $\gamma\ll c$ limit.
The processes described by the jump operators\cite{daley,breuer,ashidareview} $d_q$ in $\mathcal{D}_\downarrow$ and $d^+_q$ in $\mathcal{D}_\uparrow$ are responsible for annihilating and creating
%relaxing and exciting 
bosonic quasi-particles, respectively. Qualitatively similar results are expected for various combinations of these jump operators.
In the fermionic language, these involve energy exchange with the environment and describe a mixture of
relaxation and excitation of fermions with momentum $q$, as discussed in Ref. \cite{lindbladLL}.
The prefactors of dissipators include the Bose-Einstein distribution,
$n(T,q) = (e^{c |q|/T}-1)^{-1}$ with $T$ the temperature, ensuring that the dissipative dynamics obeys detailed balance\cite{rajagopal}~\footnote{Indeed, the ratio of the couplings of relaxation and excitation is $(1+n(T,q))/n(T,q) = e^{c |q|/T}$.}.
In addition to their physical relevance, the jump operators\cite{breuer,reichental,abbruzzo2021,brenes} are also
 motivated by the possibility of studying the
interplay between  unitary and non-unitary dynamics exactly,
by highlighting features that do not depend qualitatively on the form of the coupling to the environment.

%In addition to their physical relevance, the jump operators\cite{breuer,reichental,abbruzzo2021,brenes} are also
% motivated by the possibility of studying the 
%interplay between the unitary evolution 
%and thermalizing dissipative dynamics exactly, 
%by highlighting features that do not depend qualitatively on the form of the coupling to the environment.

%Note that the jump processes considered in Eq. \eqref{eq:jumps} are one of the many possible choices to 
%model detailed balance\cite{breuer,reichental,abbruzzo2021}, more complicated jump processes between different momentum states could also be taken into account. 
%In this work, however, we focus on the simplest but physically still relevant dissipation.

Based on the Lindbladian dynamics, the time-dependence of the following expectation values, which play crucial role in most physical quantities as we shall see later, are obtained analytically as
\begin{subequations}
\begin{gather}
n_q(t) = \Tr[\rho(t)d_q^+d_q] = n(T,q) + \left(n_0 - n(T,q) \right)e^{-\gamma q t}, \label{eq:nq}\\
m_q(t) = \Tr[\rho(t)d_q^+d_{-q}^+] = m_0 e^{(2ic - \gamma)|q|t} \label{eq:mq},
\end{gather}
\end{subequations}
where the initial values are $n_0 = (1-K)^2/(4K)$ and $m_0 = (1-K^2)/(4K)$ with $K=\sqrt{(v_F+g_4-g_2)/(v_F+g_4+g_2)}$ the LL parameter. Note that $K> 1$ and $K<1$ correspond to attractive and repulsive interaction, respectively. Furthermore, $n_q(t)$ is invariant under $K\leftrightarrow 1/K$ while 
$m_q(t)$ changes sign. 
This symmetry ensures that the our results in the present paper remain also invariant for $K\leftrightarrow 1/K$,  hence we focus on repulsive interactions 
($K<1$) only.

%In terms of $b$ bosons, we obtain
%\begin{gather}
%n_q^b(t)=\mathrm{Tr}[\rho(t)b_q^+b_q] = \frac{1}{2}\left(K + \frac{1}{K}\right) n_q(t) + \frac{1}{2}\left(K - \frac{1}{K}\right) \mathrm{Re}\left(m_q(t)\right) + \frac{\left(1-K\right)^2}{4K} \nonumber \\
%m_q^b(t)=\mathrm{Tr}[\rho(t)b_q^+b_{-q}^+] = \frac{1}{2}\left(K + \frac{1}{K}\right) \mathrm{Re}\left(m_q(t)\right) + i \mathrm{Im}\left(m_q(t)\right)  + \frac{1}{2}\left(K - \frac{1}{K}\right) \left(n_q(t) + \frac{1}{2}\right)
%\label{eq:nbmb}
%\end{gather}

\paragraph{Entropy.}
To follow the thermalization dynamics, we investigate the time-dependence of the von Neumann entropy, 
defined as $S(t) = \Tr[\rho(t)\ln\rho(t)]$. Initially, the system is in the (pure) ground state of the non-interacting Hamiltonian and, hence, $S(0)=0$. 
%As a consequence of the interaction quench, bosonic excitations are generated after the sudden quench. In addition, due to Lindbladian dynamics\cite{alba2017}, 
%the system will not remain in a pure state but will evolve into a density matrix, allowing us to inspect the interplay
%of sudden quantum quench, leaving the system in a pure state and dissipative quench, thermalizing the system.
On general ground, we expect two distinct type of time dependences. Qualitatively, after the sudden quantum quench, a significant number of excitation are created,
%especially at low energies, 
contributing to an initial sharp increase of the entropy. When the final temperature is small, these excitations need to be extracted from the
system by the dissipative process, thus after a maximum, the entropy should decrease. As we show below, after reaching a minimum, the entropy increases to its 
steady
state value.
On the other hand, for large final temperatures, even more excitations are present in the thermal state than those created by the sudden quantum quench, therefore
the entropy is expected to continue its initial increase monotonically throughout the time dependence.

The full time dependent entropy is evaluated from 
\begin{gather}
S(t) = 2\sum_{q>0}\left[(N_q(t)+1)\ln(N_q(t)+1)-N_q(t) \ln N_q(t)\right],
\label{eq:ent}
\end{gather}
where
%\begin{gather}
$N_q(t) = \sqrt{\left(n_q(t) + \frac{1}{2}\right)^2 - |m_q(t)|^2} - \frac{1}{2}\,$.
%\label{eq:Nq}
%\end{gather}
Its behaviour is best understood by inspecting the time dependence of $N_q(t)$ for 
several wavenumbers\cite{EPAPS}. 
For large $q$, when $ q> 2\ell_T^{-1} \ln|(1+K)/(1-K)|$ with the thermal length $\ell_T=c/T$, the initial ramp up period, corresponding to 
generating bosons, ends at a maximum value and
the $N_q(t)$ decreases to its steady value which is set by the Bose-Einstein distribution $n(T,q)$. This feature is a consequence of that 
the quantum quench generates more bosons 
than the expected value in thermal equilibrium and, hence, the dissipative coupling to environment extracts the surplus of bosons from the
 system and causes thermalization. 
For small wavenumbers, however, the thermal equilibrium value of bosons is large since $n(T,q)\sim T/q$ and, therefore, 
the number of bosons keeps increasing during thermalization after the initial boson generation. 

\begin{figure}[h]
\centering
\includegraphics[width=8cm]{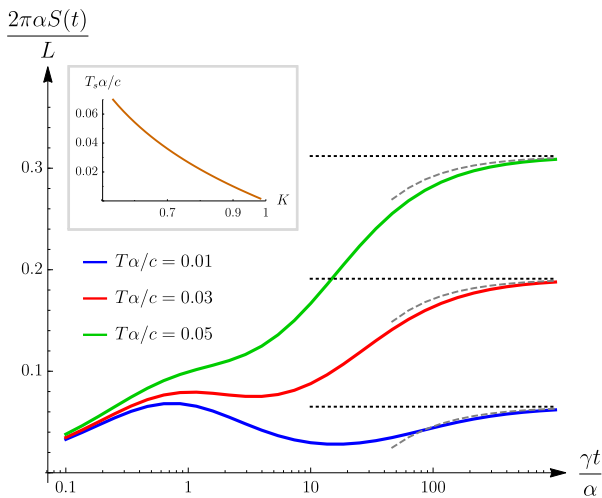}
\caption{The time evolution of the entropy at several temperatures and $K=0.7$. The black dotted line indicates the steady, thermal value. 
The gray, dashed line exhibit Eq. \eqref{eq:entlate} describing the late time behavior of the entropy. Inset: $K$-dependence of the temperature $T_{s}$ separating the two characteristics of the entropy evolution.}
\label{fig:ent}
\end{figure}

The time evolution of the entropy is shown in Fig. \ref{fig:ent} for several values of the temperature
using the conventional momentum space cut-off\cite{giamarchi} as $\exp(-\alpha q)$ with $\alpha$ the short distance cutoff. 
Initially, it grows as $t\ln(t)$ and exhibits two distinct typical behaviors for longer times, depending on the environmental temperature, as discussed above.

At low temperature, the entropy first reaches a maximum and then decreases due to the extraction of bosons out of the higher momentum modes. 
This decaying regime ends at a minimum and  is followed by a slowly growing part of the function which leads to the steady 
value of the entropy, $S_{th}(T)=L\pi T/3c$, which is the thermal entropy as
\begin{gather}
%S(t) = S_\infty + \frac{L}{\pi}\left(\frac{n_0 c}{T\left( \gamma t\right)^2} - \frac{1}{\gamma t} \right)
S(t) = S_{th}(T) - \frac{L}{\pi\gamma t},
\label{eq:entlate}
\end{gather}
which is valid for $\gamma t\gg \ell_T$. At zero temperature, however, this condition can never be fulfilled and, in this special case, the entropy 
 decreases  all the way to its thermal value which is zero.
On the other hand, at high temperatures, the entropy increases monotonically and also reaches its thermal value as in Eq. \eqref{eq:entlate}. 
This feature is the consequence of the short thermal length which dictates monotonically increasing number of bosons for most wavenumbers. 

The two characteristic time evolution is separated by the temperature $T_{s}$ which depends only on $K$ as shown in the inset of Fig. \ref{fig:ent}, 
but is independent of $\gamma/c$ and $\alpha$ for $\alpha \ll \ell_T$ ~\footnote{We note that qualitatively similar features are observed for other momentum space 
regularization schemes, e.g. sharp cutoff. The value of $T_s$ may depend on the regularization but its existence and 
the presence of the two different characteristic time evolution seem to be universal. } .
This universal feature is a consequence of the interplay between the initial boson generation by the quantum quench and the thermalization driven by 
the coupling to the environment. 
Similar behavior can be observed in the time evolution of other thermodynamic 
quantities, such as the R\'enyi entropy or total energy \cite{EPAPS}. 

\paragraph{Single-particle density matrix.}
Correlation functions display more complex behaviour than the entropy. To gain insight into the time evolution of spatial correlations, we study the equal-time single-particle density matrix.
Using the fermionic field operator $\Psi_R(x)$ \cite{giamarchi}, the single-particle density matrix is defined as $G(x,t)=\Tr\left[\rho(t)\Psi_R(x)\Psi_R(0)^+\right]$, yielding\cite{lindbladLL}
\begin{gather}
G(x,t) = G_0(x)\exp\left(-\sum_{q>0}\frac{4\pi}{Lq}(1-\cos(qx)) n_q^b(t)\right)\,.
\label{gexact}
\end{gather}
where $G_0(x)=1/(2\pi(x+i\alpha))$ is the initial correlation function obeying the well-known $1/x$ decay of free fermionic correlations\cite{giamarchi}.
The expectation value of non-interacting $b$-bosons, $n_q^b(t) = \Tr[\rho(t) b_q^+b_q]$, is 
\begin{gather}
n_q^b(t)=\frac{\left(1-K\right)^2}{4K} + \frac{K^2+1}{2K} n_q(t) + \frac{K^2-1}{2K} \mathrm{Re}\left(m_q(t)\right)\,.
\label{eq:nbq}
\end{gather}
From these, the spatio-temporal dependence of the single-particle density matrix is calculated exactly\cite{EPAPS}. 
Since the resulting expression is not too illuminating, we focus on its behaviour in various limits of interest.
We also work in the scaling limit, when $|x|$, $2ct$ and $\ell_T$ (and their combinations)  are all much larger than $\alpha$.
However, in order to make contact with the non-dissipative sudden quantum quench 
results\cite{Cazalilla2016,cazalillaprl,Iucci_Cazalilla_LLquench_PRA2009}, 
we allow for  $\gamma t$ to be comparable to $\alpha$. As already advertised, we use the weak coupling for the Lindblad equation, $c\gg \gamma$.

We start by investigating the steady state, when thermalization occurs. 
From $n_q^b(\infty)=\frac{\left(1-K\right)^2}{4K} + \frac{K^2+1}{2K}n(T,q)$, we obtain the characteristic thermal LL behaviour\cite{giamarchi,cazalillaboson} as
\begin{gather}
G^{LL}(x,\infty) =\frac{1}{2\pi\alpha}
\left\{\begin{array}{ll}
\bigl({\textstyle \frac{\alpha}{x}}\bigr)^{\scriptstyle(K+K^{-1})/2}, &  x \ll \ell_T \,,\\
\left(\frac{2\pi \alpha }{\ell_T}\right)^{\frac{K+K^{-1}}{2}}e^{-\frac{\pi(K+K^{-1})}{2}\frac{x}{\ell_T}}, &   \ell_T \ll x\,,
\end{array}\right.
\label{eq:steady}
\end{gather}
which remains also valid for the region $x\ll\gamma t$. For distances shorter than the thermal length, it follows the characteristic  power-law decay with the conventional LL exponent. 
For long distances, exponential decay is found as expected in thermal equilibrium \cite{giamarchi}.

%Between the initial and steady states, the time evolution of the single-particle density matrix is calculated by using the exponential cut-off. 
%The Green's function is plotted in Fig. \ref{fig:gK0.7}. 
%The complete, analytical result can be found in the Supplementary Material. It is, however, physically more meaningful to study the spatial decay in limiting cases.

%\begin{figure*}[h]
%\centering
%\includegraphics[width=18cm]{goftK0.7.eps}
%\caption{The single-particle density matrix.}
%\label{fig:gK0.7}
%\end{figure*}

\begin{figure}[h]
\centering
\includegraphics[width=8cm]{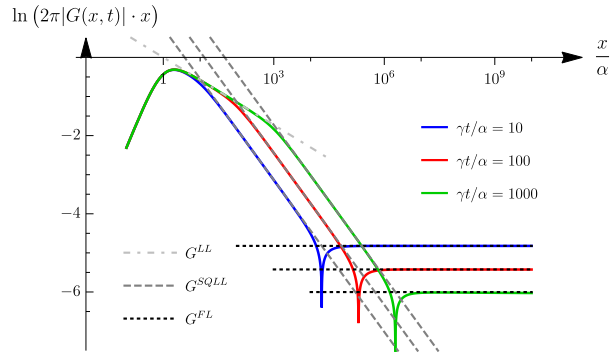}
\caption{Spatial decay of the single-particle density matrix at several time instants in the early stage of time evolution for 
$\gamma t\ll \ell_T$ with $K=0.5$, 
$c/\gamma=10^3$ and $\alpha/\ell_T=10^{-5}$. The three characteristic behaviours, namely thermal LL, dissipation enhanced sudden quench LL and dissipation enhanced Fermi liquid
behaviours are clearly visible with increasing $x$.
}
\label{fig:gearly}
\end{figure}

We now focus on the early stage of time evolution, i.e., when $\alpha + \gamma t\ll \ell_T$ but $\gamma t\ll x$, and find
\begin{gather}
%G(x,t) = G_0(x)\left(\frac{\alpha}{\mathrm{min}(x;2ct)}\right)^{-\left(\frac{K-K^{-1}}{2}\right)^2} \times \nonumber \\
%\times \left(\frac{\alpha + \gamma t}{\alpha}\right)^{\frac{(1-K)^2(K^2+1)}{4K^2}}\,.
\frac{G(x,t)}{G_0(x)}=\left(\frac{\alpha}{\mathrm{min}(x,2ct)}\right)^{-\left(\frac{K-K^{-1}}{2}\right)^2}\left(1+\frac{\gamma
t}{\alpha}\right)^{\frac{(1-K)^2(K^2+1)}{4K^2}}\,.
\label{desqll}
\end{gather}
The first term represent the conventional sudden quench result\cite{Iucci_Cazalilla_LLquench_PRA2009,cazalillaprl}, while the second one is responsible for a \emph{dissipation induced enhancement} factor, which is always bigger than 1.
This enhancement is understood by realizing that after a non-dissipative quantum quench, correlations start to propagate along light cones\cite{calabresecardy} and suppress the single-particle density matrix (i.e. the first term in Eq. \eqref{desqll}). 
%However, due to coupling to dissipative environment, this propagation is slowed down significantly due to the quantum Zeno effect\cite{misra,barontini,ashidareview}. 
However, this propagation is slowed down significantly due to the quantum Zeno effect induced by the continuously  coupling to environment.
As a result, the suppression of correlation by the sudden quench is not as effective as without the dissipative environment, therefore
an additional enhancement factor due to dissipation appears.

It is illuminating to further analyze Eq. \eqref{desqll} in limiting cases. For $\alpha+\gamma t\ll 2ct \ll x$, the single-particle density matrix 
displays dissipation enhanced Fermi liquid behaviour, namely it retains its initial spatial decay as $G^{FL}(x,t)=G_0(x)Z(t)\sim Z(t)/x$, characteristic to a Fermi liquid with time dependent Landau's quasiparticle weight\cite{cazalillaprl} as
\begin{gather}
Z(t)=\left(\frac{2ct}{\alpha}\right)^{-\left(\frac{K-K^{-1}}{2}\right)^2} \left(1+\frac{\gamma t}{\alpha}\right)^{\frac{(1-K)^2(1+K^2)}{4K^2}}\,.
\label{eq:earlytimeZ}
\end{gather}
In the absence of dissipation ($\gamma=0$), we recover previous results\cite{Iucci_Cazalilla_LLquench_PRA2009,cazalillaprl} 
as $Z(t)\sim t^{-((K-K^{-1})/2)^2}$. In the presence of dissipation, on the other hand,
the temporal decay in the quasiparticle weight is less pronounced due to the dissipation induced enhancement 
factor as $Z(t)\sim  t^{-(1-K)^2/(2K)}$. The absolute value of this second exponent in $Z(t)$ for 
$\gamma>0$ is always smaller than the one corresponding to 
$\gamma=0$. %due to the quantum Zeno effect.% as discussed above.

For $\alpha + \gamma t\ll x \ll 2ct$, we find dissipation enhanced sudden quench LL behaviour from Eq. \eqref{desqll} as
\begin{gather}
G^{SQLL}(x,t)=\frac{1}{2\pi\alpha} \left(\frac{\alpha}{x}\right)^{\left(\frac{K+K^{-1}}{2}\right)^2} \left(1+\frac{\gamma 
t}{\alpha}\right)^{\frac{(1-K)^2(K^2+1)}{4K^2}},
\label{sqll}
\end{gather}
describing the same spatial decay as found for a sudden quantum quench in Ref. \cite{Iucci_Cazalilla_LLquench_PRA2009}, featuring the sudden quench LL exponent.
The correlations are also enhanced in time by dissipation in this regime.% for $\gamma>0$.
The spatial decay of the single-particle density matrix for early times is shown in Fig. \ref{fig:gearly}, reflecting the presence of the limiting cases.

\begin{figure}[h]
\centering
\includegraphics[width=8cm]{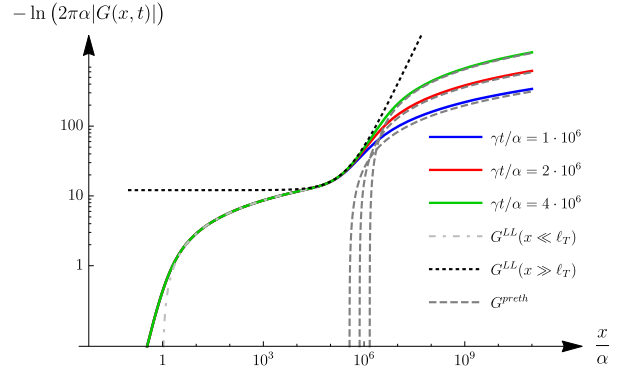}
\caption{Spatial decay of the single-particle density matrix at several time instants for late times, 
$\gamma t\gg \ell_T$ with  $K=0.5$, $c/\gamma=10^3$ and $\alpha/\ell_T=10^{-5}$. The three main characteristics, 
namely power-law decay within the thermal correlation length, thermal exponential decay above the 
thermal correlation length and 
prethermal LL behaviour are identified with increasing $x$. %{\color{red} Jelolesekre vigyazni kell, nincs kulon GLL es G thermal definialva sehol. Definaltam GLL-t, ami a teljes thermalt jelenti, ezt lenne erdemes mutatni ezen az abran inkabb} 
}
\label{fig:glate}
\end{figure}

For the late time evolution with $\ell_T\ll \gamma t \ll x$, we find peculiar behaviour what we coin as prethermal LL response as
\begin{gather}
G^{\mathrm{preth}}(x,t) = \frac{1}{2\pi\ell_T}
%\left(\frac{\ell_T}{x}\right)^{(K+K^{-1})\frac{\gamma t}{\ell_T}}\left(\frac{\gamma t}{e\ell_T}\right)^{(K+K^{-1})\frac{\gamma t}{\ell_T}}
\left(\frac{\gamma t}{x e}\right)^{(K+K^{-1})\frac{\gamma t}{\ell_T}},
\label{prethermal}
\end{gather}
following a special power-law spatial decay with a time-dependent exponent as well as exponential temporal decay. 
However, the exponent itself can be arbitrarily large, and causes a very rapid suppression of correlations~\footnote{Eq. (\ref{prethermal}) is also
multiplied by additional power-law terms as in Eq. (\ref{sqll}), but the dominant decay stems from the time dependent exponent. Neglecting these
 power-laws is further justified in Fig. \ref{fig:glate}, where the exact expression from Eq. (\ref{gexact}) is compared to Eq. (\ref{prethermal}).}. 
In the opposite, $x \ll \gamma t$ region, the steady state behavior is dominated by the thermal LL physics from Eq. \eqref{eq:steady}. 
The features of late time behavior can be recognized in Fig. \ref{fig:glate}.

\begin{figure}[h]
\centering
\includegraphics[width=8cm]{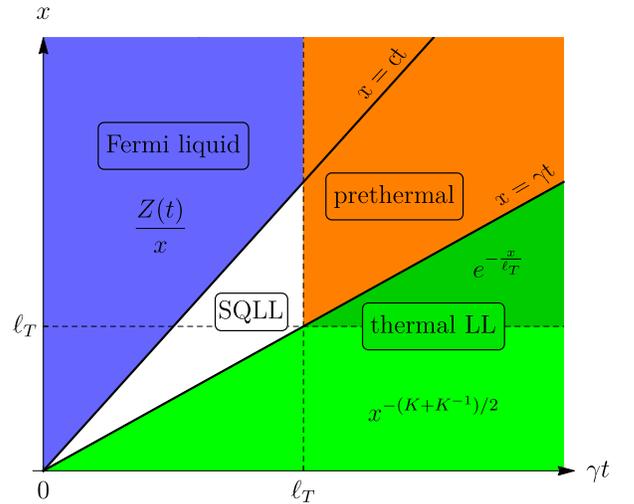}
\caption{The spatio-temporal map or "phase diagram" of the single particle density matrix, displaying the various behaviours such as dissipation enhanced sudden quench LL, dissipation enhanced Fermi liquid, prethermal and thermal LL.}
\label{fig:map}
\end{figure}

Based on these results, we can construct the spatio-temporal map or "phase diagram" of the single particle density matrix, revealing the various regions induced by the intricate interplay of unitary and non-unitary dynamics from sudden quantum quench and dissipation quench, respectively. This is shown in Fig. \ref{fig:map}.
Finally, we mention that in the absence of dissipation ($\gamma=0$), the line of $x=\gamma t$ is tilted to zero and vanishes from the map, together with $\ell_T\rightarrow\infty$. Consequently, we are only left with the regions of the Fermi liquid behavior and the sudden quench LL behavior (SQLL) without the dissipation induced enhancement.

\paragraph{Summary.}
We study thermalization and the interplay of unitary and non-unitary dynamics on an interacting one-dimensional system, a Luttinger liquid. 
By using a Lindblad equation description with couplings to environment satisfying detailed balance, thermalization is guaranteed in the long time limit.
Before reaching it, the von Neumann entropy of the system, which quantifies the amount of entanglement with the environment, either displays a maximum and minimum during the time evolution or grows
monotonically before hitting its thermal value. 
These features are expected to be generic for a wide range of systems when undergoing simultaneous quantum and environmental quench since these are direct consequences of the initial creation of excitations and the subsequent dissipative dynamics.

Correlation functions exhibit even more complex behaviour, what we demonstrate by focusing on the fermionic single particle density matrix.
We find that the presence of dissipation can enhance sudden quench correlations through the dissipation induced Zeno dynamics. 
We identify several distinct behaviours during the space-time evolution corresponding to  thermal, prethermal and dissipation enhanced sudden
quench Luttinger liquid behaviour as well as dissipation enhanced Fermi liquid response.
Our results apply not only to fermions but to bosons and spins as well.

\begin{acknowledgments}
This research is supported by the National Research, Development and
Innovation Office - NKFIH  within the Quantum Technology National Excellence
Program (Project No.~2017-1.2.1-NKP-2017-00001), K134437, K142179 and by the BME-Nanotechnology
FIKP grant (BME FIKP-NAT), and by a grant of the Ministry of Research, Innovation and
 Digitization, CNCS/CCCDI-UEFISCDI, under projects number PN-III-P4-ID-PCE-2020-0277.
\end{acknowledgments}

%%%%%%%%%%%%%%%%%%%%%%%%%%%%%%%%%%%%%%%%

\bibliographystyle{apsrev}
\bibliography{lindblad,wboson1}

\appendix

\section{Time evolution of the total energy and R\'enyi entropy}
Similarly to the entropy presented in the main text, other thermodynamic quantities also feature two different time-dependence depending on the temperature of the thermalizing environment. In this Supplementary Material, we demonstrate the time evolution of the total energy and the R\' enyi entropy. The latter is defined as $E(t) = \Tr[\rho(t)H]$ with $H$ the Hamiltonian of the interacting system.

\begin{figure}[h]
\centering
\includegraphics[width=8cm]{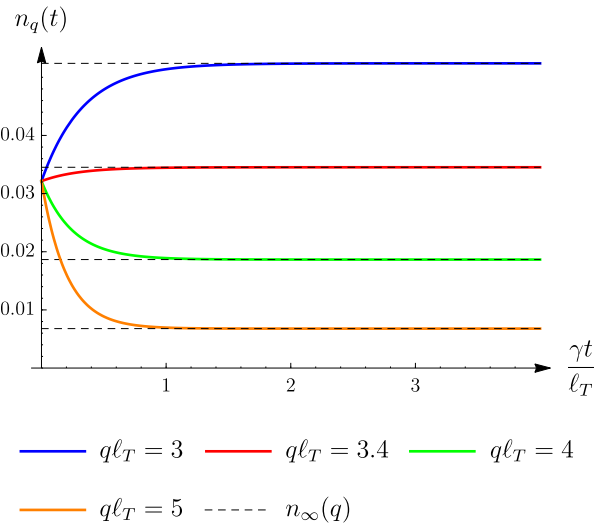}
\caption{Time-dependence of $n_q(t)$ at different wavenumbers. The function starts at $n_0(K) = (1-K)^2/(4K)$ in each wavenumber channel and then monotonously tends to the thermal value, $n(T,q)$.}
\label{fig:nqt}
\end{figure}

The total energy is evaluated as
\begin{gather}
E(t) = E_0 + 2\sum_{q>0}cqn_q(t)
\label{eq:tote}
\end{gather}
where $E_0 = \sum_{q>0} (c-v_F-g_4)q$ is the interacting ground state energy. The time-dependence occurs through the quantity of $n_q(t)$ defined in 
Eq. \eqref{eq:nq} of the main text, respectively. Fig. \ref{fig:nqt} shows that $n_q(t)$ starts at the same initial value, $n_0=(1-K)^2/4K$ and approaches exponentially the steady value, $n(T,q)$.

\begin{figure}[h]
\centering
\includegraphics[width=9cm]{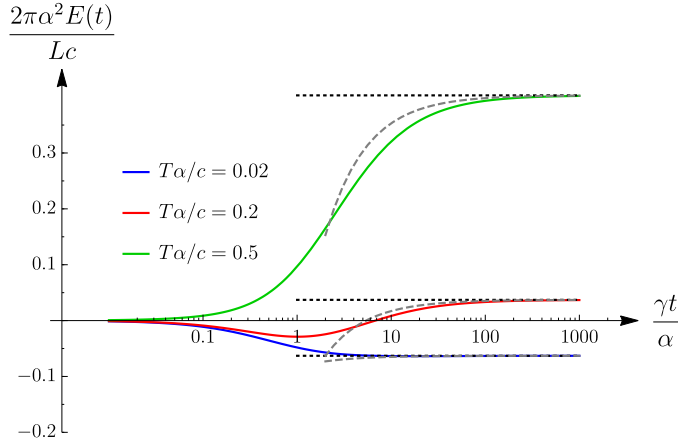}
\caption{The time evolution of total energy at several temperature values. The black dotted lines indicate the steady value of the energy, \eqref{eq:einf}. The gray dashed lines show the long time behavior, \eqref{eq:elt}. For the plot, $K=0.7$ is set.}
\label{fig:tote}
\end{figure}

Substituting \eqref{eq:nq} into \eqref{eq:tote}, the integral over $q$ yields
\begin{gather}
E(t) = \frac{Lc}{2\pi\alpha^2}\left[-\frac{(1+K)^2}{2K} \frac{\gamma t(2\alpha + \gamma t)}{(\alpha + \gamma t)^2} +  \right. \nonumber \\ 
\left. +
2\left( \frac{\alpha}{\ell_T} \right)^2 \left( \Psi_1\left( \frac{\alpha + \gamma t}{\ell_T}\right) - \Psi_1\left( \frac{\alpha}{\ell_T}\right) \right) \right]
\end{gather}
with $\Psi_1(y)$ the polygamma function. The long time behavior of the total energy is obtained as
\begin{gather}
E(t) = E_{th}(T) -\frac{Lc}{2\pi\ell_T\gamma t}
\label{eq:elt}
\end{gather}
with the steady state thermal value
\begin{gather}
E_{th}(T) = - \frac{Lc}{2\pi\alpha^2}\left(\frac{(1+K)^2}{2K} + 2\left( \frac{\alpha}{\ell_T} \right)^2 \Psi_1\left( \frac{\alpha}{\ell_T}\right)\right)\,.
\label{eq:einf}
\end{gather}
The complete time evolution is shown in Fig. \ref{fig:tote}. It can be seen that at high temperatures, the total energy monotonously increases, indicating that the initially generated amount of bosons is less than that of the thermal state. At low temperatures, however, an initial reduction can be observed due to the extraction of bosons on the high momentum (high energy) modes. Later, the energy increases back when high energy modes are already thermalized and only low energy modes exhibit dynamics.

\begin{figure}[h]
\centering
\includegraphics[width=8cm]{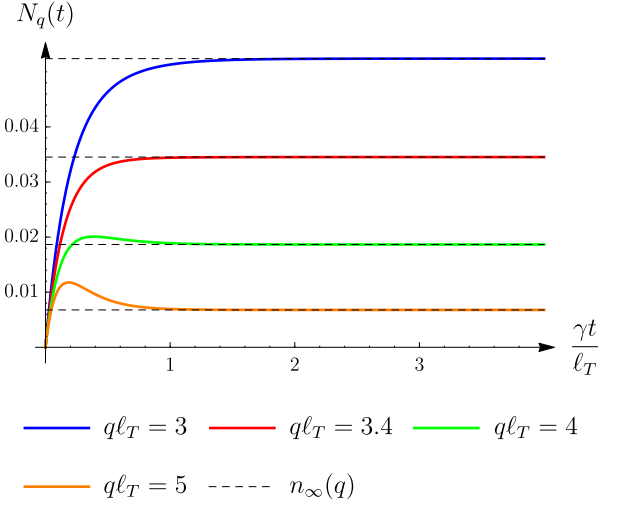}
\caption{Time-dependence of $N_q(t)$ at different wavenumbers. At high wavenumbers, $N_q(t)$ exhibits a maximum and decreases back to its thermal value, $n(T,q)$. At low wavenumbers, the function increases monotonously. Furthermore, it is remarkable that high momentum modes thermalize faster than low energy modes.}
\label{fig:Nqt}
\end{figure}

Among thermodynamic quantities, the R\' enyi-2 entropy plays an important role because of its experimental accessibility \cite{Islam2015,Brydges}.  The R\' enyi entropy is defined as $S_2(t)=-\ln\Tr\left[\rho(t)^2\right]$ and is calculated as
\begin{gather}
S_2(t) = 2\sum_{q>0} \ln\left(2N_q(t) +1\right)
\end{gather}
with $N_q(t)$ defined below Eq. \eqref{eq:ent}. Fig. \ref{fig:Nqt} shows that in the low momentum (low energy) channels, $N_q(t)$ increases to its thermal value 
monotonously while at high $q$, the function has a maximum after which it decreases to the thermal value.

\begin{figure}[h]
\centering
\includegraphics[width=9cm]{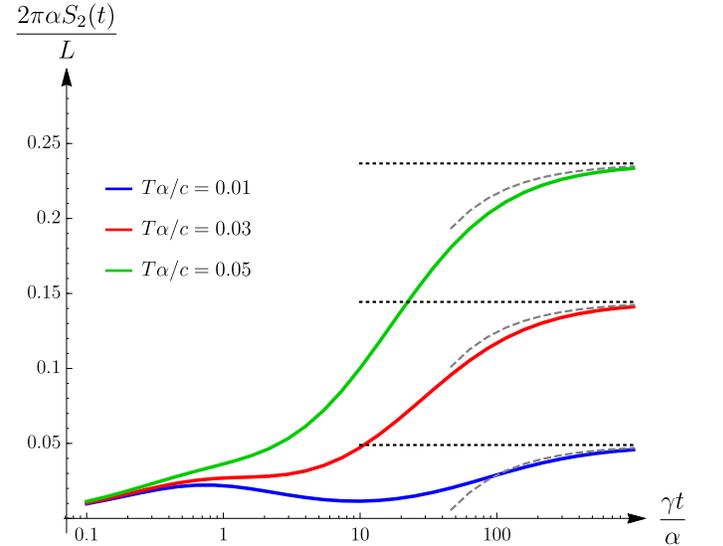}
\caption{The time evolution of the R\'enyi entropy at several temperature values. The black dotted lines indicate the steady state thermal value, $S_{2,T}$. The gray dashed lines show the long time behavior which is analytically obtained as $S_2(t)=S_{2,T} - L/(\pi\gamma t)$. For the plot, $K=0.7$ is set.}
\label{fig:renyi}
\end{figure}

The R\'enyi entropy is computed numerically and the typical time evolution scenarios are shown in Fig. \ref{fig:renyi}. At high temperatures, the entropy increases monotonically while at low temperatures, the initial growth of the entropy is followed by a decreasing period. 

To summarize, similarly to the von Neumann entropy presented in the main text, both the total energy and the R\'enyi entropy exhibit two characteristic time evolution. These features can be explained by the interplay between the quantum quench and the dissipative dynamics. The quantum quench generates bosons in each momentum channels. For low momenta, this amount of boson is still less than it should be at thermal equilibrium and, hence, the dissipative dynamics will further increase the amount of bosons. For higher momentum, the quench-generated is already more than it should in equilibrium and, therefore, dissipation extracts the surplus. By changing the temperature, the boarder between low and high momentum regimes are shifted resulting in temperature dependent characteristics in thermodynamic quantities.

\section{Analytic derivation of the equal-time single-particle density matrix}
As presented in the main text, the single particle density matrix is obtained as
\begin{gather}
G(x,t) = G_0(x)\exp\left(-\sum_{q>0}\frac{4\pi}{Lq}(1-\cos(qx)) n_q^b(t)\right)\,.
\label{eq:gr1}
\end{gather}
where $G_0(x)$ is the initial Green's function and $n^b_q(t)$ is the average number of $b$-bosons calculated in Eq. \eqref{eq:nbq} of the main text. We substitute \eqref{eq:nbq} into \eqref{eq:gr1} and take the thermodynamic limit leading to a $q$-integral instead of the sum. To regularize the integral, we use an exponential cutoff, $\sum_q \rightarrow \int\mathrm{d}q\,e^{-\alpha q}$. The integral is carried out as
\begin{widetext}
\begin{gather}
\ln\left(\frac{G(x,t)}{G_0(x)}\right) = 
\frac{K^2+1}{2K} \left[\ln\left(\frac{\Gamma\left(1 + \frac{\alpha + \gamma t}{\ell_T}\right)^2}{\Gamma\left(1 + \frac{\alpha + \gamma t + ix}{\ell_T}\right)\Gamma\left(1 + \frac{\alpha + \gamma t-ix}{\ell_T}\right)}  \right) - \ln\left(\frac{\Gamma\left(1 + \frac{\alpha}{\ell_T}\right)^2}{\Gamma\left(1 + \frac{\alpha + ix}{\ell_T}\right)\Gamma\left(1 + \frac{\alpha - ix}{\ell_T}\right)}  \right)   \right] + \nonumber \\ 
+ \frac{(1-K^2)^2}{16K^2}\ln\left(\frac{\left((\alpha+\gamma t)^2 + \left(x-2ct\right)^2  \right)\left((\alpha+\gamma t)^2 + \left(x+2ct\right)^2  \right)}{\left((\alpha+\gamma t)^2 + \left(2ct\right)^2  \right)^2}\right) - \nonumber \\
- \frac{(K^2 + 1)(1-K)^2}{8K^2}\ln\left(1 + \frac{x^2  }{(\alpha+\gamma t)^2}\right) - \frac{(1-K)^2}{4K}\ln\left(1+\left(\frac{x}{\alpha}\right)^2\right)
\end{gather}
\end{widetext}
where $\Gamma(y)$ is the Gamma function.

\end{document}